\documentclass[12pt]{article}
\pdfoutput=1 
\usepackage{geometry}
\usepackage[title]{appendix}
\usepackage[authoryear,round]{natbib}
\usepackage[T1]{fontenc}
\usepackage{braket}
\usepackage{setspace}
\usepackage{titlesec}
\usepackage[dvipsnames]{xcolor}
\definecolor{acmilan}{HTML}{b52e2b}
\definecolor{mediumpersianblue}{rgb}{0.0, 0.4, 0.65}
\definecolor{pblue}{rgb}{0.11, 0.22, 0.73}
\usepackage{hyperref}

\usepackage{amsmath,amsthm,amssymb,ifthen,mathrsfs,amsfonts}

\usepackage{empheq}
\usepackage{mathtools}
\usepackage{graphicx}
\usepackage{accents}
\usepackage{enumitem}
\usepackage{bigints}
\usepackage{tikz}

\usetikzlibrary{arrows,intersections}

\newtheorem{theorem}{Theorem}
\newtheorem{lemma}[theorem]{Lemma}
\newtheorem{proposition}[theorem]{Proposition}
\theoremstyle{definition}

\newtheorem{definition}[theorem]{Definition}

\newtheorem{notation}[theorem]{Notation}

\newtheorem{example}[theorem]{Example}

\newtheorem{assumption}[theorem]{Assumption}

\newcommand{\BE}{\mathsf{E}}
\newcommand{\BP}{\mathsf{P}}

\newcommand{\BR}{\mathbb{R}}
\newcommand{\BN}{\mathbb{N}}

\newcommand{\mc}[1]{\mathcal{#1}}

\renewcommand{\d}{\mathrm{d}}
\newcommand{\ve}{\varepsilon}
\newcommand{\Var}{\mathsf{Var}}
\newcommand{\absl}[1]{\left| #1\right |}
\DeclarePairedDelimiter{\norm}{\lVert}{\rVert}

\newcommand{\cond}{\left| \right.}

\numberwithin{equation}{section}

\makeatletter

\newcommand{\Rn}[1]{\expandafter\@slowromancap\romannumeral #1@}
\makeatother

\title{High Dimensional Decision Making, Upper and Lower Bounds}
\date{May 2021}
\author{Farzad Pourbabaee \thanks{Email: \href{mailto:farzad@berkeley.edu}{farzad@berkeley.edu} \newline Consortium for Data Analytics in Risk, Department of Economics, 530 Evans Hall \#3880, University of California, Berkeley, CA 94720-3880.}}
\renewcommand\footnotemark{}
\begin{document}
\maketitle
\begin{abstract}
    A decision maker's utility depends on her action $a\in A \subset \BR^d$ and the payoff relevant state of the world $\theta\in \Theta$. One can define the \textit{value} of acquiring new information as the difference between the maximum expected utility pre- and post information acquisition. In this paper, I find asymptotic results on the expected value of information as $d \to \infty$, by using tools from the theory of (sub)-Guassian processes and generic chaining.
\end{abstract}

\noindent {\it JEL classification:} C44; D81

\noindent {\it Keywords:} Decision making; Information valuation; High dimensional vectors

\clearpage
\section{Setup}
In many economic environments the decision making process entails several dimensions, e.g allocation of a divisible good to a \textit{large} group of consumers with heterogeneous tastes, or the consumer's problem when the number of available goods is \textit{large}, and hence the consumption portfolio is a high-dimensional vector. Other contexts where the decision making environment could potentially be high dimensional includes the contextual bandit \cite{bastani2020online}\footnote{As well as the ongoing literature on Bandits and exploration/exploitation trade-offs in complex environments, see e.g \cite{li2019k} and \cite{pourbabaee2020robust}.}, incomplete contracts with undescribable contingencies \cite{al2006undescribable} and model selection and inference in complex environments \cite{al2014coarse}.

Conceptually, let $u(a,\theta)$ be the utility of a decision maker (DM) where $a \in A$ is her action and $\theta \in \Theta$ is the payoff-relevant state of the world, hidden to the DM. The action space $A$ is a compact subset of a high-dimensional vector space say $\BR^d$ (where $d$ is large), and $u$ is continuous in $a$.

In the absence of any information about $\theta$, the DM takes an action $a$ that maximizes the expected utility $\bar{v}_a := \BE u(a,\theta)$. Now, suppose there is an information source, denoted by the random variable $S$ conditionally distributed given $\theta$, that is available to the DM. Therefore, she maximizes her conditional expected utility, i.e $v_a(S):= \BE\left[ u(a,\theta) \cond S\right]$, given the signal $S$. The \textit{realized value} associated with the signal $S$ is the difference between the DM's maximum expected utility pre- and post signal realization, that is
\begin{equation*}
	\sup_{a \in A}\BE\left[\left.u(a,\theta) \right| S\right]-\sup_{a \in A} \BE\left[u(a,\theta)\right]=\sup_{a \in A}v_a(S)- \sup_{a \in A} \bar{v}_a.
\end{equation*}
Thus, one defines the \textit{expected value of information}\footnote{For an axiomatic treatment of the cost (or value) of information see the recent work by \cite{pomatto2020cost}.} by
\begin{equation}
	\mc{V}:= \BE \left[\sup_{a \in A}v_a(S)- \sup_{a \in A} \bar{v}_a\right].
\end{equation}
Defining the \textit{centered} random variable $\tilde{v}_a(S):=v_a(S)-\bar{v}_a$, it readily follows
\begin{equation*}
	\mc{V} \leq \widetilde{\mc{V}}:=\BE\left[\sup_{a \in A} \tilde{v}_a(S)\right],
\end{equation*}
simply because $\sup_{a \in A}v_a(S)- \sup_{a \in A} \bar{v}_a \leq \sup_{a \in A} \tilde{v}_a(S)$. If the DM is indifferent between all actions in $A$ before the signal realization, namely $\bar{v}_a = \bar{v}$ for all $a \in A$, then the above inequality binds and in that case she is called an \textit{indifferent} DM.

Given the signal structure $S \cond \theta$, the goal of this paper is to find sharp upper and lower bounds for $\widetilde{\mc{V}}$, that translate to bounding the expected information value $\mc{V}$ for an indifferent DM. The leading tool that is used throughout is the \textit{Generic Chaining} developed in \cite{talagrand2006generic} and \cite{talagrand2014upper}.

\section{Generic chaining}
In this section, I present some definitions related to Generic Chaining and Talagrand's seminal  approach that paves the way for presenting the upper and lower bounds. 
\begin{definition}[Intrinsic metric]
\label{def: intrinsic_metric}
	For every $a,a' \in A$, define $\rho(a,a') := \BE\left[\left(\tilde{v}_a(S)-\tilde{v}_{a'}(S)\right)^2\right]^{1/2}$. Then, $\rho: A \times A \to \BR_+$ induces a pseudometric on the action space $A$, that is called the \textit{intrinsic metric}. The \textit{diameter} of this space is $\text{diam}(A):=\sup_{a,a' \in A}\rho(a,a')$.
\end{definition}
\begin{definition}[Increment condition]
\label{assum: increment_cond}
	The \textit{stochastic process} $\tilde{v}:=\left\{\tilde{v}_a(S):a \in A\right\}$ satisfies the \textit{increment condition} if
	\begin{equation}
	\label{eq: increment_cond}
		\BP\left(\absl{\tilde{v}_a(S)-\tilde{v}_{a'}(S)} \geq t \right) \leq 2\exp\left(-\frac{t^2}{2\rho(a,a')^2}\right), ~ \forall a,a' \in A, \text{ and } t \in \BR_+.
	\end{equation}
	Equivalently, this means the process $\tilde{v}$ has \textit{sub-Gaussian} increments.
\end{definition}
\begin{definition}[Admissible sequence and Talagrand's functional]
	Given the action space $A$, an \textit{admissible sequence} is an \textit{increasing} sequence $\left\{\mc{A}_n: n\in \BN\right\}$ of partitions of $A$, such that $\absl{\mc{A}_n} \leq 2^{2^n}$ for $n\geq 1$ and $\absl{\mc{A}_0}=1$. It is called \textit{increasing} in the sense that every member of $\mc{A}_{n+1}$ is a subset of an element of $\mc{A}_n$.\footnote{More precisley, let $\mc{A}_n=\left\{A_{n}^i: 1\leq i \leq m_n\right\}$ be a partition for $A$. The sequence of partitions $\left\{\mc{A}_n: n\in \BN\right\}$ is called increasing if for every $A_{n+1}^i\in \mc{A}_{n+1}$, there exists $A_n^j \in \mc{A}_n$ such that $A_{n+1}^i \subseteq A_n^j$.} For each $a \in A$, let $\mc{A}_n(a)$ be the corresponding set in $\mc{A}_n$ that contains $a$, then the Talagrand's functional is defined as
	\begin{equation}
		\gamma_2(A,\rho) := \inf \sup_{a \in A} \sum_{n\geq 0} 2^{n/2} \text{diam}\left(\mc{A}_n(a)\right),
	\end{equation}
	where the infimum is taken over are all admissible sequences for the set $A$.
\end{definition}
\begin{notation}[Inequalities upto universal constants] We say $f \lesssim g$ (or $f \gtrsim g$) if there exits a universal constant $c\in \BR_{++}$, particularly independent of the dimension $d$, such that $f \leq c g$ (or $f \geq c g$). Similarly, $f \simeq g$ if $f \lesssim g$ and $f \gtrsim g$.
\end{notation}
We are now ready to state the application of Talagrand's upper bound to our setup that follows directly from theorem 2.2.22 of \cite{talagrand2014upper}.
\begin{theorem}[Sharp upper bound] Assume for the intrinsic metric space $\left(A,\rho\right)$ the process $\tilde{v}$ satisfies the increment condition \ref{eq: increment_cond}, then
	\begin{equation}
		\mc{V} \leq \widetilde{\mc{V}} \lesssim \gamma_2 (A,\rho).
	\end{equation}
\end{theorem}
The above theorem holds for all decision making environments and signal structures in which $\tilde{v}$ satisfies the increment condition. That is with almost minimal structure on the primitives of the problem, an upper bound on the value of information $S$ is obtained, which as it turns out in the next theorem is quite sharp in the \textit{Gaussian} case. More specifically, next theorem argues when the DM is indifferent (so $\mc{V}=\widetilde{\mc{V}}$) and the information structure is Gaussian (so $\widetilde{\mc{V}} \simeq \gamma_2(A,\rho)$), the asymptotic value of information is \textit{completely} characterized by Talagrand's functional (i.e $\mc{V}=\gamma_2(A,\rho)$). The following result is the Majorizing Measure Theorem (\cite{talagrand2014upper} theorem 2.4.1) applied to our setup.
\begin{theorem}[Matching lower bound for Gaussians]
\label{thm: match_lower_bound}
	If the process $\left\{\tilde{v}_a(S):a \in A\right\}$ is jointly Gaussian, then $\mc{\widetilde{V}} \simeq \gamma_2(A,\rho)$, namely there are universal constants $(c,C) \in \BR^2_{++}$ such that
	\begin{equation}
		c \gamma_2(A,\rho) \leq \mc{\widetilde{V}} \leq C \gamma_2(A,\rho).
	\end{equation}
\end{theorem}
The substantive content of this theorem is that it determines the value of $\mc{\widetilde{V}}$ up to universal constants, that are in particular independent of the dimension $d$.

\section{Demystifying the functional}
To understand the functional $\gamma_2(A,\rho)$, it is imperative to figure out the pseudometric $\rho$. To simplify the picture, assume for now that the DM is indifferent, therefore before the signal realization she values all her choices $a \in A$ similarly. The essential role that $S$ plays is to \textit{separate} the actions from each other by assigning the ex-post value $v_a(S)$ to each action $a \in A$, hence the \textit{distance} between two actions $a,a' \in A$ is measured by $\rho(a,a')=\BE\left[\left(v_a(S)-v_{a'}(S)\right)^2\right]^{1/2}$.\footnote{Since the DM is assumed indifferent here, then $v_a(S)-v_{a'}(S) = \tilde{v}_a(S)-\tilde{v}_{a'}(S)$.} To build intuition, one measure of how good the signal $S$ (or equivalently the metric $\rho$) carries out this separation is the diameter of set $A$, i.e $\text{diam}(A)$. As a result of this, on could \textit{loosely} say the larger the diameter of the action space (induced by the intrinsic metric $\rho$), the more \textit{valuable} is the signal $S$. However, this is just a first-order type of refinement, that is the notion of diameter is too coarse to measure the worth of information contained in $S$. A fine-tuning thus needed by zooming in on the set $A$ recursively, through an admissible sequence of partitions, and that is exactly what $\gamma_2(A,\rho)$ does. 

Unfortunately, in most cases direct computation of the functional $\gamma_2(A,\rho)$ becomes far too complicated. In such instances, one approach is to work out upper and lower bounds for $\gamma_2(A,\rho)$. Occasionally, the upper and lower bounds match, thereby indirectly pinning down $\gamma_2(A,\rho)$. For this purpose we need some preliminary concepts mostly developed in the statistical learning literature and can be found in \cite{vershynin2018high}.
\begin{definition}[Covering and Packing number]
	For the metric space $(A,\rho)$ and given $\ve>0$, the covering number is the minimum number of $\ve$-balls $\mathbb{B}_\rho(\ve)$ required to cover $A$, which is denoted by $N(A,\rho,\ve)$. The packing number $M(A,\rho,\ve)$ is the maximum number of points in $A$ that their pairwise distance is larger than $\ve$.
\end{definition}
Two standard results in this vein that relate the covering and packing number together are stated in the next lemma. The proofs are in the appendix.
\begin{lemma}
\label{lem: cov-pack}
	Given the metric space $(A,\rho)$,
	\begin{enumerate}[label=(\alph*)]
		\item \label{item: cov-pack-a} for every $\ve>0$:
			\begin{equation}
				M(A,\rho,2 \ve) \leq N(A,\rho,\ve) \leq  M(A,\rho,\ve).
			\end{equation}
		\item \label{item: cov-pack-b} Let $\mathbb{B}^d_\rho$ be the $d$-dimensional unit ball in $(A,\rho)$. Then, for every subset $D\subset A$
			\begin{equation}
			\label{eq: cov-pack-b}
				\frac{\textsf{Vol}\left(D\right)}{\textsf{Vol}\left(\ve \mathbb{B}^d_\rho\right)} \leq N(D,\rho,\ve) \leq \frac{\textsf{Vol}\left(D+\frac{\ve}{2}\mathbb{B}^d_\rho\right)}{\textsf{Vol}\left(\frac{\ve}{2}\mathbb{B}^d_\rho\right)},
			\end{equation}
			where $\textsf{Vol}$ denotes the Euclidean volume.
	\end{enumerate}
\end{lemma}
Having defined the notion of covering number, I can now state an important technique to upper and lower bound $\gamma_2(A,\rho)$.
\begin{lemma}
\label{lem: approx_gamma}
	In the metric space $(A,\rho)$, the Talagrand's functional is bounded by 
	\begin{equation}
	    \label{eq: dudley}
		\ve \sqrt{\log N(A,\rho,\ve)} \lesssim \gamma_2(A,\rho) \lesssim \int_0^{\text{diam}(A)}\sqrt{\log N(A,\rho,\ve)}\d \ve, 
	\end{equation}
	where the lower bound holds for every $\ve>0$.\footnote{The integral in the \textit{rhs} of \eqref{eq: dudley} is called the Dudley's entropy integral developed in \cite{dudley1967sizes}.}
\end{lemma}
As I will show in the next section, there are cases where the above bounds match, thus providing sharp characterization for $\gamma_2(A,\rho)$ and hence for the expected value of information $\mc{V}$.

\section{Gaussian environment}
In this section we restrict our attention to the case where the payoff-relevant state of the world is a high-dimensional Gaussian vector $\theta \sim \mc{N}(0,\Sigma_\theta)$, and the diagonal elements of $\Sigma_\theta$ are normalized to one. In the Gaussian environment we can use the matching lower bound presented in theorem \ref{thm: match_lower_bound}, and hope to find sharp expressions for the expected value of information rather than just differential upper and lower bounds. In this section, let the DM's utility function be $u(a,\theta) = \langle a, \theta \rangle$ on the compact subset $A \subset \BR^d$. By construction, such a DM is indifferent, so any upper and lower bound on $\widetilde{\mc{V}}$ also hold for $\mc{V}$. Finally, assume the signal structure is Gaussian, namely
\begin{equation}
\label{eq: gaussian_signal}
	(\theta,S)^T \sim \mc{N}\left(0, \begin{bmatrix}
		\Sigma_\theta & \Sigma_{\theta s}\\
		\Sigma_{s \theta} & \Sigma_s
	\end{bmatrix}\right).
\end{equation}
Define $W:= \Sigma_{\theta s}\Sigma_s^{-1}\Sigma_{s \theta}$, then the intrinsic metric follows
\begin{equation*}
	\begin{split}
	 	\rho(a,a')^2 &= \BE\left[ \langle a-a', \BE\left[\theta | S\right] \rangle ^2 \right]\\
	 	&=\BE\left[(a-a')^T \Sigma_{\theta s}\Sigma_{s}^{-1}SS^T\Sigma_{s}^{-1}\Sigma_{s \theta}(a-a')\right]=(a-a')^T W(a-a').
	\end{split}
\end{equation*}
Therefore, the metric $\rho$ is induced by the norm $\norm{a}_W:=\sqrt{a^T W a}$. The unit ball under the metric $\rho$, or equivalently under the norm $\norm{\cdot}_W$, is a $d$-dimensional ellipsoid $\mathbb{B}^d_W=\mathbb{B}^d_\rho=\left\{a\in \BR^d: a^T W a \leq 1\right\}$, and its volume is equal to $\frac{\pi^d}{\Gamma(\frac{d}{2}+1)}\det(W)^{-1/2}$. In the following, I assume the matrix $W$ that shapes the conditional variance, i.e $\Var(\theta \cond S)$, is invertible. 
\begin{example}
\label{exp: perfect_info_benchmark}
As a benchmark let $A = \mathbb{B}^d_\infty = \left[-1,1\right]^d$, and assume the signal $S$ is perfectly conclusive (i.e $S=\theta$), thus the maximum expected value of information is
\begin{equation}
	\mc{V}^* = \BE\left[\sup_{a \in \mathbb{B}^d_\infty} \langle a ,\theta \rangle \right]= \BE \norm{\theta}_1 = d \BE \absl{\theta_1}=d \sqrt{2/\pi},
\end{equation}
where $\norm{\cdot}_1$ is the $\ell_1$ norm.
Therefore, the maximum value that an expert can charge for providing information $S$ is proportional to $d$, when the action space is the $\ell_\infty$ unit ball.
\end{example} 
In the next proposition, I show when $A\in \left\{\mathbb{B}^d_\infty, \mathbb{B}^d_2\right\}$, that are respectively the $\ell_\infty$ and $\ell_2$ unit balls, the lower and upper bounds offered for $\gamma_2(A,\rho)$ in lemma \ref{lem: approx_gamma} match, and hence we get sharp characterization for the value of information $S$.  

Recall that $W$ is the covariance matrix of $\theta$ given $S$, that is positive semidefinite. Let $\lambda_{(1)} \leq \ldots \leq \lambda_{(d)}$ be the set of its eigenvalues. Further assume the information acquisition technology is \textit{limited}, in a sense that there exists an irreducible component in $\theta$ that is never resolved by $S$, so $\lambda_{(1)} \geq \underline{\lambda} >0$. In addition, since $\text{Var}\left(\theta | S\right) \preceq \Sigma_\theta$, there is a natural upper bound $\lambda_{(d)}\leq \bar{\lambda}$. The matrix $W$ is called \textit{$(\underline{\lambda},\bar{\lambda})$-bounded} if $\underline{\lambda} \leq \lambda_{(1)} \leq \lambda_{(d)}\leq \bar{\lambda}$.
\begin{assumption}
It is assumed in the following that the pair $(\underline{\lambda},\bar{\lambda})$ are independent of the dimension $d$. That is as the dimension of the covariance matrix increases, the range of its eigenvalues remains bounded.
\end{assumption}

\begin{proposition}
\label{prop: budget_set_effect}
	If the covariance matrix $W$ remains $(\underline{\lambda},\bar{\lambda})$ bounded as the dimension $d$ grows, then $\mc{V}\left(A = \mathbb{B}^d_2\right) \simeq \sqrt{d}$ and $\mc{V}\left(A = \mathbb{B}^d_\infty\right) \simeq d$.
\end{proposition}
The interesting result of the above proposition is that one knows how the value of acquiring more information changes with respect to the dimension of the decision making environment. The proof of this result (presented in the appendix) owes to the fact that for $A\in \left\{\mathbb{B}^d_\infty, \mathbb{B}^d_2\right\}$ the upper and lower bounds in lemma \ref{lem: cov-pack} match each other (up to universal constants) and hence a sharp characterization for $\mc{V}$ is obtained.

\section{Concluding remarks}
This paper explores a novel method to analyze the value associated with acquiring information in high dimensional settings. It is shown that the informational value of a signal, defined as the expected difference between the maximum utility pre- and post signal realization, can be summarized in terms of the \textit{geometric shape} of the action space $A$ through the lens of a functional that takes in $A$ and a pseudometric $\rho$ expressing the separation power of the signal. Some examples are offered, where one can compute \textit{sharp} characterization of this functional when the dimension of the action space grows.

\appendix
\section{Proofs}
\subsection{Proof of lemma \ref{lem: cov-pack}}
	For the upper bound in part \ref{item: cov-pack-a}, suppose one finds an $\ve$-packing $\{a_1,\ldots, a_M\}$ such that the distance between every two points of this set is larger than $\ve$. Now if the collection of $\ve$-balls around these points do not cover the entire $A$, then a point $a_0 \in A$ can be found that is not contained in this collection and hence is at least $\ve$ away from all these $M$ points, so $M$ is not maximal and can be enlarged by adding $a_0$.
	
	For the lower bound in part \ref{item: cov-pack-a}, suppose one finds a $2\ve$-packing $\{a_1,\ldots,a_M\}$ and an $\ve$-covering with centers at $\{b_1,\ldots,b_N\}$ such that $M\geq N+1$. Therefore, there must exist two elements of the packing set, say $a_1$ and $a_2$, that are located in a same $\ve$-ball around the centers of the covering set, thus their distance must be less than or equal to $2\ve$ that violates the packing condition. 
	
	For the lower bound in part \ref{item: cov-pack-b}, let $\left\{a_1,\ldots,a_N\right\}$ be an $\ve$-covering for $D$, so $D \subset \bigcup_{i=1}^N \left(a_i+\ve \mathbb{B}^d_\rho\right)$, and therefore $\textsf{Vol}\left(D\right) \leq N \textsf{Vol}\left(\ve \mathbb{B}^d_\rho\right)$, yielding the lower bound.
	
	For the upper bound in part \ref{item: cov-pack-b}, let $\left\{a_1,\ldots,a_M\right\}$ be a maximal $\ve$-packing for $D$, therefore the collection of $\ve/2$-balls, $\left\{a_i+\frac{\ve}{2}\mathbb{B}^d_\rho: 1\leq i \leq M\right\}$, are disjoint from each other and their union is a subset of $D + \frac{\ve}{2}\mathbb{B}^d_\rho$, so
	\begin{equation*}
		M \textsf{Vol}\left(\frac{\ve}{2}\mathbb{B}^d_\rho\right) \leq \textsf{Vol}\left( D+ \frac{\ve}{2}\mathbb{B}^d_\rho\right),
	\end{equation*}
	that together with $N(D,\rho,\ve) \leq M$ imply the upper bound in \eqref{eq: cov-pack-b}.\qed

\subsection{Proof of lemma \ref{lem: approx_gamma}}
	For the proof of upper bound refer to section 2.2 of \cite{talagrand2014upper}.
	
	To prove the lower bound, assume $\ve>0$ is given. Then, let $k = \min \left\{n: N(A,\rho,\ve) \leq 2^{2^n}\right\}$. For every admissible sequence of partitions $\left\{\mc{A}_n\right\}$, and every $a \in A$,
	\begin{equation*}
		\sum_{n\geq 0}2^{n/2} \text{diam}\left(\mc{A}_n(a)\right) \geq 2^{(k-1)/2} \text{diam}\left(\mc{A}_{k-1}(a)\right)
	\end{equation*}
	hence expressing the partition $\mc{A}_{k-1}:=\left\{A^{(i)}_{k-1}: 1\leq i \leq 2^{2^{k-1}}\right\}$ implies 
	\begin{equation*}
		\begin{split}
			\sup_{a \in A} \sum_{n\geq 0}2^{n/2} \text{diam}\left(\mc{A}_n(a)\right) &\geq 2^{(k-1)/2} \sup_{a \in A} \text{diam}\left(\mc{A}_{k-1}(a)\right)\\
			&=2^{(k-1)/2} \max\left\{\text{diam}\left(A_{k-1}^{(i)}\right):1\leq i \leq 2^{2^{k-1}}\right\}\\
			&\geq \frac{1}{\sqrt{2}}\sqrt{\log N(A,\rho,\ve)} \max\left\{\text{diam}\left(A_{k-1}^{(i)}\right):1\leq i \leq 2^{2^{k-1}}\right\}.
		\end{split}
	\end{equation*} 
	Note that $\delta:=\max\left\{\text{diam}\left(A_{k-1}^{(i)}\right):1\leq i \leq 2^{2^{k-1}}\right\}\geq \ve$, because otherwise if $\delta < \ve$ one can choose $a^{(i)}_{k-1}\in A^{(i)}_{k-1}$ for each $i \leq 2^{2^{k-1}}$ such that the collection of $\delta$-balls around these points constitute a $\delta$-cover for $A$, and hence $N(A,\rho,\delta) \leq 2^{2^{k-1}}$, that in turn means $N(A,\rho,\ve) \leq 2^{2^{k-1}}$. This contradicts the minimality of $k$. Therefore, for every admissible sequence of partitions $\left\{\mc{A}_n\right\}$ and every $\ve>0$:
	\begin{equation*}
		\sup_{a \in A} \sum_{n\geq 0}2^{n/2} \text{diam}\left(\mc{A}_n(a)\right) \geq \frac{1}{\sqrt{2}}\ve\sqrt{\log N(A,\rho,\ve)} 
	\end{equation*}
	Taking the infimum over all admissible partitions on the \textit{lhs} justifies the lower bound.\qed

\subsection{Proof of proposition \ref{prop: budget_set_effect}}
	\underline{Case $A = \mathbb{B}^d_\infty$}: A valid upper bound is $\mc{V} \lesssim d$, because no information structure can outperform the perfect information benchmark of example \ref{exp: perfect_info_benchmark}, that scales with $d$. Toward the lower bound, I first find a lower bound for $N(A,\rho,\ve)$. From the inequality in part \ref{item: cov-pack-b} of lemma \ref{lem: cov-pack} we have
	\begin{equation*}
		N(A,\rho,\ve) \geq \frac{\textsf{Vol}\left(\mathbb{B}^d_\infty\right)}{\textsf{Vol}\left(\ve \mathbb{B}^d_\rho\right)}=\left(\frac{2}{\pi \ve}\right)^d \sqrt{\det(W)}\Gamma\left(\frac{d}{2}+1\right),
	\end{equation*}
	therefore,
	\begin{equation*}
		\ve \sqrt{\log N(A,\rho,\ve)} \geq \ve \sqrt{d\log \frac{2}{\pi \ve}+ \log \sqrt{\det(W)}\Gamma\left(\frac{d}{2}+1\right)}.
	\end{equation*} 
	Set $\ve = \frac{2}{\pi \sqrt{e}} \left(\sqrt{\det(W)}\Gamma\left(\frac{d}{2}+1\right)\right)^{1/d}$ in the \textit{rhs} above. Further, let $m = [d/2]$, so $d/2 \geq m$. Application of Sterling's approximation (i.e $m! \geq m^{m+1/2}e^{-m}$) implies $(m!)^{1/2m} \gtrsim \sqrt{m}$, then
	\begin{equation*}
		\begin{split}
			\ve \sqrt{\log N(A,\rho,\ve)} &\geq \frac{\sqrt{2/e}}{\pi} \det(W)^{1/2d}\Gamma\left(\frac{d}{2}+1\right)^{1/d}\sqrt{d}\\
			&\geq \frac{\sqrt{2/e}}{\pi} \det(W)^{1/2d}(m!)^{1/2m}\sqrt{d}\\
			&\gtrsim \left[\det(W)^{1/2d}\right] d \geq \underline{\lambda}^{1/2}d.
		\end{split}
	\end{equation*}
	Therefore, using the lower bound in lemma \ref{lem: approx_gamma} implies:
	\begin{equation}
		 \sqrt{\underline{\lambda}} d \leq \left[\det(W)^{1/2d}\right] d \lesssim \mc{V}\left(\mathbb{B}^d_\infty\right)\lesssim d
	\end{equation}
	
	\noindent\underline{Case $A = \mathbb{B}^d_2$}: Again from the lower bound in part \ref{item: cov-pack-b} of lemma \ref{lem: cov-pack} we have
	\begin{equation*}
		N(A,\rho,\ve) \geq \frac{\textsf{Vol}\left(\mathbb{B}^d_2\right)}{\textsf{Vol}\left(\ve \mathbb{B}^d_\rho\right)}=\left(\frac{1}{ \ve}\right)^d \sqrt{\det(W)},
	\end{equation*}
	therefore,
	\begin{equation*}
		\ve \sqrt{\log N(A,\rho,\ve)} \geq \ve \sqrt{d\log \frac{1}{\ve}+ \log \sqrt{\det(W)}}.
	\end{equation*}
	Set $\ve = e^{-1/2}\det(W)^{1/2d}$ in the \textit{rhs}, and it is easy again to check 
	\begin{equation}
	\label{eq: lower_B_2}
	    \ve \sqrt{\log N(A,\rho,\ve)}  \gtrsim \left[\det(W)^{1/2d}\right] \sqrt{d} \geq \sqrt{\underline{\lambda}d}.
	\end{equation}
	From the upper bound in part \ref{item: cov-pack-b} of lemma \ref{lem: cov-pack}, for every $\ve>0$:
	\begin{equation*}
		N(\mathbb{B}^d_2,\rho,\ve) \leq \frac{\textsf{Vol}\left(\mathbb{B}^d_2+\frac{\ve}{2}\mathbb{B}^d_\rho\right)}{\textsf{Vol}\left(\frac{\ve}{2}\mathbb{B}^d_\rho\right)}
	\end{equation*}
	Let $\lambda_{(1)} \leq \ldots \leq \lambda_{(d)}$ be set of eigenvalues for $W$. The unit ball $\mathbb{B}_\rho^d$ is an ellipsoid in $\BR^d$ with elliptic radii $\frac{1}{\lambda_{(1)}^{1/2}} \geq \ldots \geq \frac{1}{\lambda_{(d)}^{1/2}}$. Therefore, $\mathbb{B}_2^d + \frac{\ve}{2}\mathbb{B}^d_\rho \subset \left(1+\ve/2\lambda_{(1)}^{1/2}\right)\mathbb{B}_2^d$, and as a result of this
	\begin{equation*}
		N(\mathbb{B}^d_2,\rho,\ve) \leq \left(\frac{2}{\ve}+\frac{1}{\lambda_{(1)}^{1/2}}\right)^d \sqrt{\det(W)}.
	\end{equation*}
	In addition, under the metric $\rho$ the diameter of $\mathbb{B}^d_2$ is 
	\begin{equation}
	    \sup\left\{\sqrt{(x-y)^T W (x-y)}: x^Tx \leq 1, y^Ty \leq 1 \right\},
	\end{equation}
	that can be readily confirmed to be equal to $\text{diam}(\mathbb{B}^d_2)=2\lambda_{(d)}^{1/2}$. Therefore, the Dudley's entropy integral is upper-bounded as
	\begin{equation*}
		\begin{split}
			\int_0^{\text{diam}(\mathbb{B}^d_2)} \sqrt{\log N(A,\rho,\ve)} \d \ve &\leq \sqrt{d} \int_0^{2\lambda_{(d)}^{1/2}} \sqrt{\log \left(2/\ve+1/\lambda_{(1)}^{1/2}\right)+\log\left( \det(W)^{1/2d}\right)}\d \ve \\
			&\leq  \sqrt{d} \int_0^{2\bar{\lambda}} \sqrt{\log \left(2/\ve+1/\underline{\lambda}^{1/2}\right)+\log \bar{\lambda}^{1/2}}\d \ve \lesssim \sqrt{d}.
		\end{split} 
	\end{equation*}
	The second inequality is resulted from $(\underline{\lambda},\bar{\lambda})$-boundedness, and the third one falls out because the integral is finite. Inserting the above upper bound and the lower bound of \eqref{eq: lower_B_2} in lemma \ref{lem: approx_gamma} imply that:
	\begin{equation*}
		\mc{V}\left(\mathbb{B}^d_2\right) \simeq \sqrt{d}
	\end{equation*}\qed
\setcitestyle{numbers}	 
\bibliographystyle{plainnat}
\bibliography{ref}
\end{document}